\documentclass[12pt,english]{article}
\usepackage[T1]{fontenc}
\usepackage[latin9]{inputenc}
\usepackage{geometry}
\usepackage{color}
\usepackage{babel}
\usepackage{float}
\usepackage{amsmath}
\usepackage{amssymb}
\usepackage{graphicx}
\usepackage[unicode=true,pdfusetitle,
 bookmarks=true,bookmarksnumbered=false,bookmarksopen=false,
 breaklinks=false,pdfborder={0 0 0},pdfborderstyle={},backref=false,colorlinks=true]
 {hyperref}

\makeatletter

\newcommand{\lyxdot}{.}

\@ifundefined{date}{}{\date{}}
\usepackage{lmodern}
\usepackage[T1]{fontenc}
\usepackage{caption}
\usepackage{upgreek}

\@ifundefined{showcaptionsetup}{}{%
 \PassOptionsToPackage{caption=false}{subfig}}
\usepackage{subfig}
\makeatother

\begin{document}
\title{\textcolor{black}{First results from plasma edge biasing on SPECTOR
}\\
\textcolor{black}{\large{}\vspace{1cm}
}}
\author{\textcolor{black}{Carl Dunlea$^{1*}$, General Fusion Team$^{2}$,}\\
\textcolor{black}{Chijin Xiao$^{1}$, and Akira Hirose$^{1}$}}

\maketitle
\textcolor{black}{$^{1}$University of Saskatchewan, Saskatoon, Canada }

\textcolor{black}{$^{2}$General Fusion, Vancouver, Canada}

\textcolor{black}{$^{*}$e-mail: cpd716@mail.usask.ca }

\textcolor{black}{\thispagestyle{empty} }
\begin{abstract}
\textcolor{black}{A description of an edge-biasing experiment conducted
on the SPECTOR (Spherical Compact Toroid) plasma injector is presented,
along with initiaresults. The insertion of a disc-shaped molybdenum
electrode (probe), biased at up to $+100$V, into the edge of the
CT (Compact Torus), resulted in up to 1kA radial current being drawn.
Core electron temperature, as measured with a Thomson-scattering diagnostic,
was found to increase by a factor of up to 2.4 in the optimal configuration
tested. $\mbox{H}_{\alpha}$ intensity was observed to decrease, and
CT lifetimes increased by a factor of up to 2.3. A significant reduction
in electron density was observed; this is thought to be due to the
effect of a transport barrier impeding CT fueling, where, as verified
by MHD simulation, the fueling source is neutral gas that remains
concentrated around the gas valves after CT formation. }
\end{abstract}

\section{\textcolor{black}{Introduction}}

\textcolor{black}{High confinement mode (H-mode) has been implemented
by various means ($e.g.,$ edge biasing, neutral beams, ion or electron
cyclotron heating, lower hybrid heating, and ohmic heating) on a range
of magnetic confinement configurations including tokamaks, reversed
field pinches, stellarators, and mirror machines. The first H-mode
was produced in the ASDEX tokamak by neutral beam injection in 1982
\cite{wagnerASDEX}. In 1989, H-mode was first produced by electrode
edge biasing on the CCT tokamak \cite{Taylor_CCT,WEynants_Taylor_CCT}.
In 1990, it was observed that edge impurity ion poloidal speed is
modified abruptly during transitions from low to high confinement
modes on the DIII-D tokamak \cite{Groebner}. H-mode has been produced
routinely on many magnetic-fusion experiments, including practically
all the large tokamaks, for example JET, TFTR, and JT-60. Since the
initial electrode-biasing experiments on CCT, H-mode has been produced
by edge biasing on many tokamaks, for example CASTOR \cite{CAstor,OOst},
T-10 \cite{OOst,T10}, STOR-M \cite{StorM}, ISTTOK \cite{Figueiredo_ISTTOK},
TEXTOR \cite{OOst,Jachmich}, J-TEXT \cite{JTEXT}, and TCABR \cite{TCABR}. }

\textcolor{black}{Electrode biasing involves the insertion of an electrode,
that is biased relative to the vessel wall near the point of insertion,
into the edge of a magnetized plasma. This leads to a radially directed
electric field between the probe and the wall. A commonly held interpretation
of the mechanism behind the improved confinement observed on edge
biasing experiments is that the resultant $\mathbf{J}_{r}\times\mathbf{B}$
force imposed on the plasma at the edge of the plasma confinement
region varies with distance between the probe and the wall, because
$E_{r}$, as well as the magnetic field, vary in that region. The
associated torque overcomes viscous forces, spinning up the edge plasma,
and results in shearing of the particle velocities between the probe
and the wall. The sheared velocity profile is thought to suppress
the growth of turbulent eddies that advect hot plasma particles to
the wall, thereby reducing this plasma cooling mechanism. In general,
H-modes induced by probe biasing share features of those initiated
by various methods of heating, including a density pedestal near the
wall (near the probe radius for probe biasing), diminished levels
of recycling as evidenced by reduced $\mbox{H}_{\alpha}$ emission
intensity, and increased particle and energy confinement times. For
example, increases in energy confinement times by factors of 1.5,
1.5, 1.2, 1.8 and 2  were reported for CCT, STOR-M, TEXTOR, T-10 and
TCABR  respectively. Core electron density increased by a factor of
four on CCT, while line-averaged ele ctron density increased by factors
of 2, 2, 1.5, 1.8 and 2.6 on STOR-M, TEXTOR, T-10, CASTOR and TCABR
respectively. Of these six examples, a biasing-induced temperature
increase was noted only for the T-10 experiment, with an increase
in core ion temperature by a factor of 1.4 reported, while reduced
$\mbox{H}_{\alpha}$ emission intensity was recorded in each case.
Many of these biasing experiments \cite{Taylor_CCT,WEynants_Taylor_CCT,StorM,Jachmich,TCABR}
have probed the plasma edge region, obtaining profiles of, for example,
electric field, density, temperature and/or velocity profiles, leading
to clarifications of the mechanisms involved in the transition to
H-mode. }

\textcolor{black}{Positive as well as negative electrode biasing works
well on some machines; in other instances only one biasing polarity
has the desired effect. Most biasing experiments have used passive
electrodes, while some have implemented electron-emitting electrodes.
Emissive electrodes have, in addition to a circuit to bias the electrode
relative to the vacuum vessel, a separate heating circuit, and are
heated until they emit electrons. Materials traditionally used for
emissive electrodes include lanthanum hexaboride (LaB6) and tungsten
(W). Generally speaking, emissive electrodes add complexity to an
experiment, but may be beneficial when the edge plasma electron density
is so low that dangerously high voltages (which could initiate a current
arc that could damage the electrode and vessel) would be required
in order to draw an edge current sufficiently high enough for the
$\mathbf{J}_{r}\times\mathbf{B}$ force to overcome inertial effects
(viscosity, friction) and drive edge rotation. In the CCT tokamak
\cite{Taylor_CCT}, LaB6 cathodes heated by carbon rods drew edge
current up to $40\mbox{A}$ when the voltage measured between the
electrode (probe) and vessel wall was $V_{probe}\sim-250\mbox{V}$.
On CCT, for negative bias, it was found that both electron-emissive
electrodes and passive graphite electrodes produced similar results,
as long as the electrode was large enough to draw sufficient current
($\sim20\mbox{A}$), and small enough not to form a limiter \cite{Taylor_CCT}.
For negative biasing on ISTTOK, it was not possible to draw more than
$2\mbox{ to }3\mbox{A}$ with a passive electrode, compared with  $\sim20\mbox{A}$
with an emissive electrode, while the current drawn with positive
biasing was the same for emissive and non-emissive electrode ($I_{probe}\sim28\mbox{A}$
at $V_{probe}\sim+130\mbox{V}$). }

\textcolor{black}{It is widely accepted that radial electric fields
at the edge plasma, and mitigation of turbulent transport, are involved
in the transition between L and H modes in tokamaks. However, an issue
that has been widely considered is whether it is the direct stabilization
of a turbulence-inducing plasma instability, by enhanced velocity
shear in the transport barrier, or actual eddy decorrelation and direct
energy transport reduction at the barrier, that causes the confinement
improvements that are routinely observed in H-modes. H-modes produced
by electrode biasing support the theory that it is actual eddy decorrelation
that leads to the radial transport reduction. In these cases, the
experimenter controls $E_{r}$, the radial component of the electric
field, and velocity shear directly. However, in a magnetized plasma,
stabilization of a range of modes can occur if $\omega_{E\times B}$,
the $E\times B$ shearing rate, is greater than $\omega_{max}$, the
maximum linear growth rate of the mode \cite{Tendler,Burrell}. Interaction
between the unstable mode with a nearby stable mode can lead to enhanced
Landau damping of the unstable mode \cite{Tendler}. Nonlinear gyro-Landau
fluid simulations indicate turbulence stabilization in cases where
$\omega_{E\times B}$ is comparable to $\omega_{max}$, the maximum
linear growth rate of all the unstable modes in the plasma \cite{waltz}.
According to \cite{Tendler}, for a significant reduction of turbulence,
it is not a strict requirement that $\omega_{E\times B}>\omega_{max}$
for all unstable modes in the plasma, just that $\omega_{E\times B}$
must be at least of the order of the nonlinear saturation of the turbulence
level due to all modes. Note that $\omega_{E\times B}$ enters into
the various theories quadratically \cite{Burrell}, so that its sign
irrelevant. H-mode has been observed experimentally with both signs
of $E_{r}$ and $\omega_{E\times B}$. The Ion Temperature Gradient
(ITG) mode is thought to contribute to turbulence initiation and thermal
transport in magnetically confined plasmas, in particular in plasmas
with high ion temperatures and suffciently weak density profiles \cite{Cowley }.
Nonlinear numerical analysis has shown that the ITG mode is completely
suppressed when $\omega_{E\times B}\sim\omega_{max}$ \cite{Tendler}.}

\textcolor{black}{For decorrelation of turbulence due to $E\times B$
shear, $\omega_{E\times B}$ must be of the order of $\Delta\omega_{T}$,
the turbulence decorrelation rate associated with turbulent radial
diffusion of fluctuations by the ambient turbulence \cite{Tendler,Burrell}.
$\Delta\omega_{T}$ can be calculated in some cases but it is not
generally available for comparison with experiment. It has been shown
by theoretical analysis that, contrary to expectation, a radial electric
field without dependence on the radial coordinate, can actually result
in a differential rotation ($i.e.,$ sheared $E\times B$ velocity)
as a result of geometric effects \cite{SHAING}. In that work, by
studying the development and saturation of resistivity-gradient-driven
turbulence in the presence of a radial electric field in cylindrical
geometry, the authors concluded that the turbulence decorrelation
rate was not modified when $\omega_{E\times B}\lesssim\Delta\omega_{T}$.
In a consequent publication by a different group \cite{Biglari},
it was shown analytically that coupling between poloidal shearing
and turbulent radial scattering can account for turbulence quenching
without invoking model-dependent, symmetry breaking effects such as
diamagnetic rotation. The researchers showed that if an $E_{r}$ (of
either sign) with sufficient radial shear (in either direction), can
be established one way or another, that this would lead to turbulence
suppression. They emphasized that it is a combination of radial scattering
and poloidal shearing, with a weighting on poloidal shearing, that
determines the turbulence decorrelation process. They suggest that,
since the radial decorrelation length for several turbulence models
scales inversely with magnetic field, that it is more difficult to
suppress turbulence when the magnetic field is high. Theory predicts
that an adequate $E\times B$ shearing rate results in reduction of
radial, fluctuation-driven transport. Supporting this, plasma edge
Langmuir probe measurements, for electrode biasing-induced H-modes
as well as for spontaneous H-modes in tokamaks and in the Wendelstein
7-AS stellarator, show large reductions in fluctuation-driven particle
flux \cite{Basse}. }

\textcolor{black}{Pre-biasing conditions of radial electric field
and an extensive range of plasma parameters play roles in determining
the beneficial polarity and the level of bias-induced plasma confinement
improvement \cite{Tendler}. A reduction of radial transport at the
edge would be beneficial for confinement not only because of reduced
outward thermal transport, but also due to reduced inward transport
of cold wall-recycled particles to the core. This latter effect is
especially relevant on small machines (such as SPECTOR, major radius
$R\sim19\mbox{cm},\text{ minor radius }\,a\sim8\mbox{cm}$, no limiter
or divertor) for which the surface area to volume ratio of the magnetically
confined plasma is large, particularly in configurations without a
limiter or divertor, where the recycling process is more important. }

\textcolor{black}{An overview of the experiment setup with a description
of the biasing electrode assembly is presented in section \ref{sec:Experiment-setup}.
Circuit analysis, leading to an estimate for the resistance of the
plasma between the electrode and flux conserver, which was useful
for optimising the circuit, is the focus of section \ref{sec:Circuit-analysis}.
Main results are presented in section \ref{sec:Main-results}. A discussion
of principal findings, conclusions and possible further improvements
to the experiment is presented in section \ref{sec:Discussion-and-conclusions}.
Results from simulations of neutral and plasma fluid interaction during
CT formation in SPECTOR geometry is presented in appendix \ref{sec:Neutral_SPECTOR}.
The simulations indicate that neutral gas, which remains concentrated
around the locations of the machine gas valves after CT formation,
can diffuse up the gun as a source of CT fueling. It may be partly
due to the effect of a transport barrier impeding the CT fueling process
that the improvements in confinement times and electron temperatures
observed with the initial edge biasing tests on SPECTOR appear especially
significant. }\\
\textcolor{black}{}\\
\textcolor{black}{}\\

\section{\textcolor{black}{Experiment setup\label{sec:Experiment-setup}}}

\textcolor{black}{}
\begin{figure}[H]
\centering{}\textcolor{black}{}\subfloat[]{

\textcolor{black}{\includegraphics[width=6cm,height=10cm]{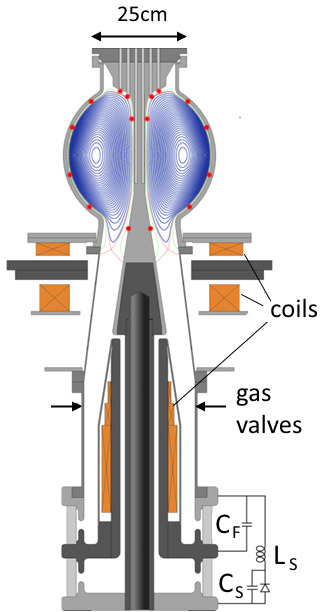}}

}\textcolor{black}{\hfill{}}\subfloat[]{\textcolor{black}{\includegraphics[width=7cm,height=5cm]{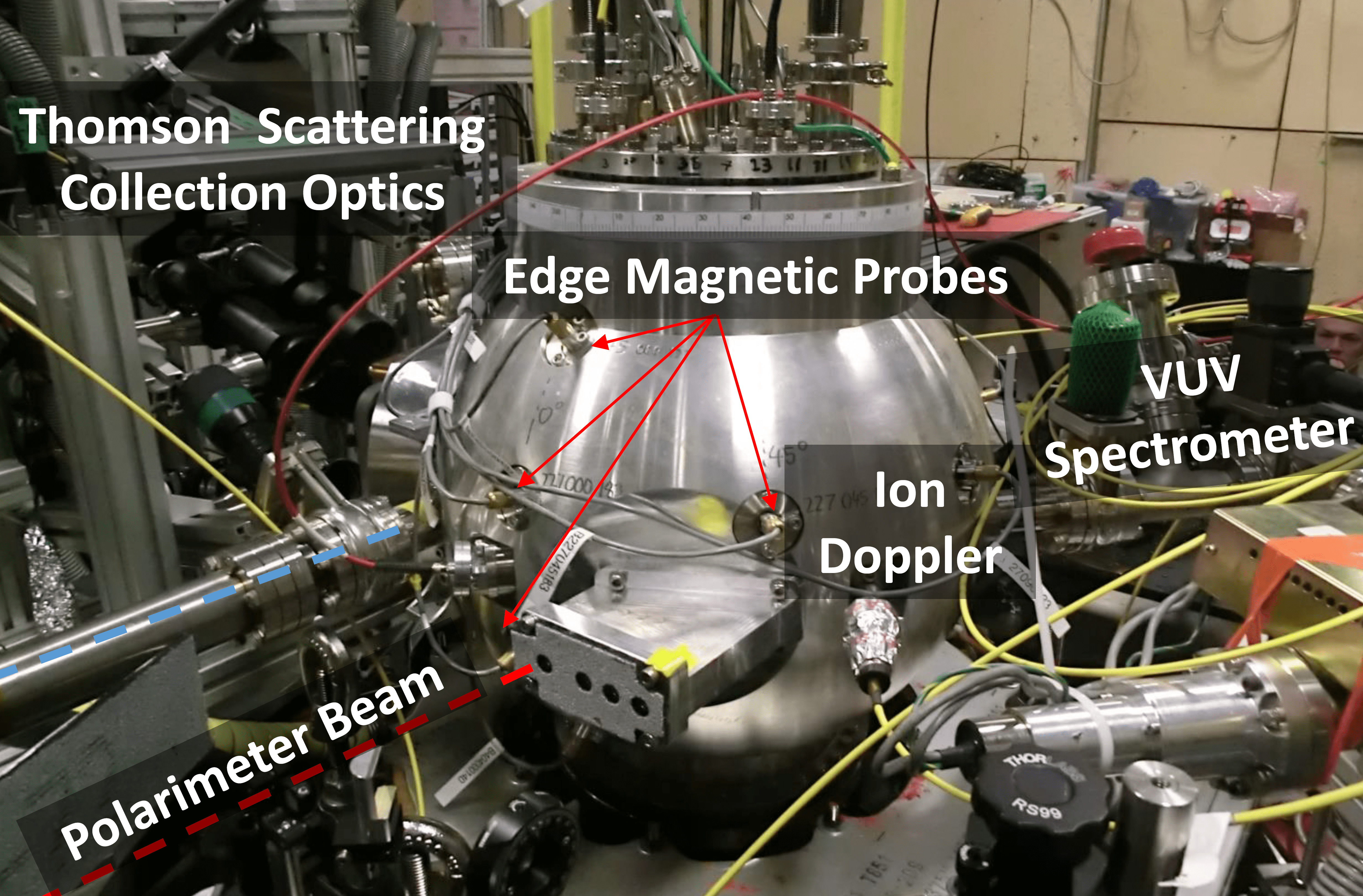}}}\textcolor{black}{\caption{\label{fig:Spector}Schematics of SPECTOR plasma injector and diagnostics{\footnotesize{} }}
}
\end{figure}
\textcolor{black}{A schematic of the SPECTOR \cite{spectPoster} plasma
injector is depicted in figure \ref{fig:Spector}(a), where the red
dots along the flux conserving wall of the CT containment region represent
the locations of magnetic probes. SPECTOR is a magnetized Marshall
gun that produces compact tori (CTs). It has, in addition to the formation
circuit that drives up to $0.8\mbox{MA}$ formation current over around
$80\upmu$s, a separate circuit to produce an approximately constant
shaft current of up to $0.5\mbox{MA}$, which flows up the outer walls
of the machine and down the central shaft, increasing CT toroidal
field and making the CT more robust against MHD instability. Shaft
current duration is extended to around 3ms with a crowbar inductor/diode
circuit, which is indicated schematically in figure \ref{fig:Spector}(a).
Toroidal field at the CT core is typically around $0.5\mbox{T}$.
The low CT aspect ratio, and the $q$ profile, define the CTs as spherical
tokamaks. Initial poloidal flux on the SPECTOR injector is up to 30
mWb. Coaxial helicity injection produces plasma currents in the range
$0.3-0.8\text{MA}$. SPECTOR has a range of plasma diagnostics including
magnetic probes, visible and VUV spectroscopy, three interferometer
chords, multi-point Thomson scattering (TS), and a four chord FIR
polarimeter system. The plasma facing components are mostly plasma-sprayed
tungsten and bare aluminum, and can be coated with lithium as a gettering
agent. Experimental investigations of formation dynamics, MHD mode
activity, evolution of plasma profiles, and machine operation setting
optimisation have been made. Safety factor profiles can be controlled
by varying shaft current and the axial distribution of poloidal gun
flux. Thomson-scattering (TS) measurements have indicated CT electron
temperatures in excess of 400eV and densities of the order 10$^{20}$
m$^{-3}$. Energy confinement times of the order of 100 ms have been
estimated \cite{spectPoster}.  A selection of TS system-produced
electron temperature and electron density measurements \cite{TS_GF}
(both taken at $300$$\upmu\mbox{s}$ after CT formation), electron
density measurements obtained with a far-infrared (FIR) interferometer
\cite{polarimetryGF}, spectral data, and magnetic probe data, will
be presented in this paper. The principal diagnostics are indicated
in figure \ref{fig:Spector}(b). }
\begin{figure}[H]
\centering{}\textcolor{black}{}\subfloat{\textcolor{black}{\includegraphics[width=14cm,height=7cm]{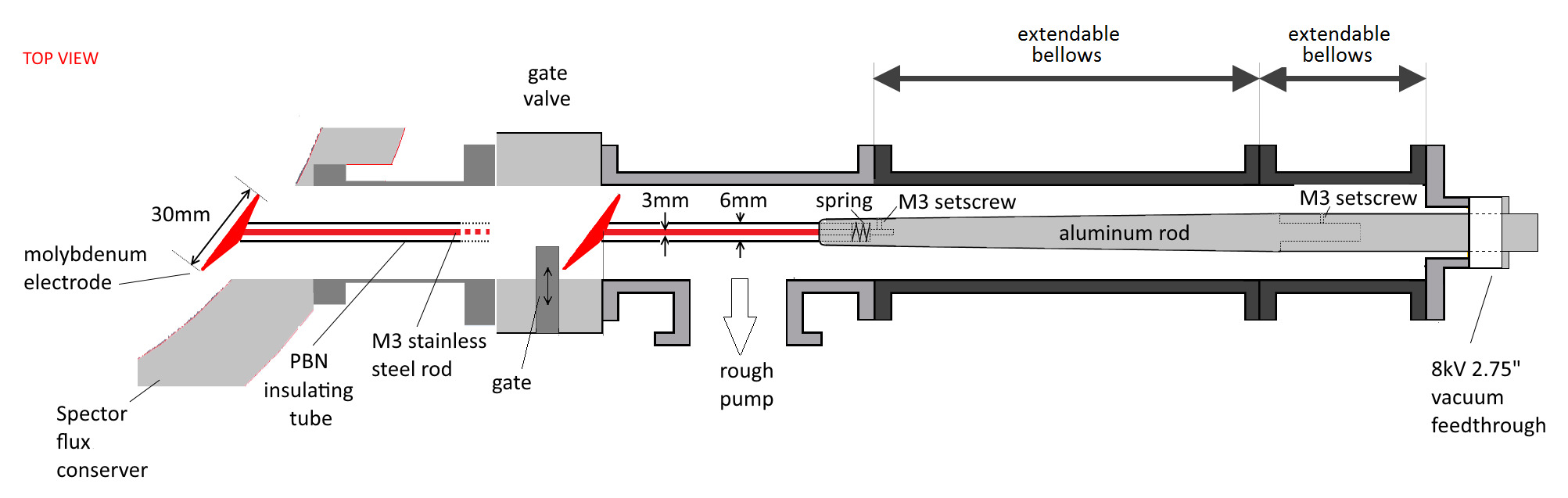}}}\textcolor{black}{\caption{\label{fig:bias probe assembly}$\,\,\,\,$Biasing probe assembly}
}
\end{figure}
\textcolor{black}{Figure \ref{fig:bias probe assembly} indicates
a top-view of the electrode (probe) assembly with extendable vacuum
bellows. The biasing electrode can be retracted behind the gate valve
and isolated from the machine vacuum. The approximately disc-shaped
electrode (tapered towards the flux-conserver facing side to enable
attachment to the steel rod, as indicated in figure \ref{fig:bias probe assembly})
is machined from molybdenum bar. It has a 30mm diameter and has an
average thickness of approximately 3mm. Molybdenum was chosen for
its high work function against sputtering, high melting point, and
its resilience against the corrosive action of lithium, which is used
as a gettering agent on SPECTOR. A pyrolytic boron nitride (PBN) tube
is used as a plasma-compatible insulator around the M3 stainless steel
rod that connects the electrode to a tapered aluminum rod, which is
in turn connected to the $0.5$ inch diameter copper rod that forms
part of the $8\mbox{kV}$ $2.75$ inch CF (ConFlat) vacuum feedthrough.
The electrode can be inserted up to 45mm into the vacuum vessel; insertion
depth was 11mm for the results presented here. Note that the probe
can be used as a limiter by connecting the vacuum feedthrough indicated
in figure \ref{fig:bias probe assembly} directly to the vacuum vessel.
This configuration was not tested because tests using a molybdenum
limiter on other GF injectors resulted in evidence of localised limiter
melting. \newpage}

\section{\textcolor{black}{Circuit analysis\label{sec:Circuit-analysis}}}

\textcolor{black}{}
\begin{figure}[H]
\centering{}\textcolor{black}{}\subfloat{\textcolor{black}{\includegraphics[width=16cm,height=7cm]{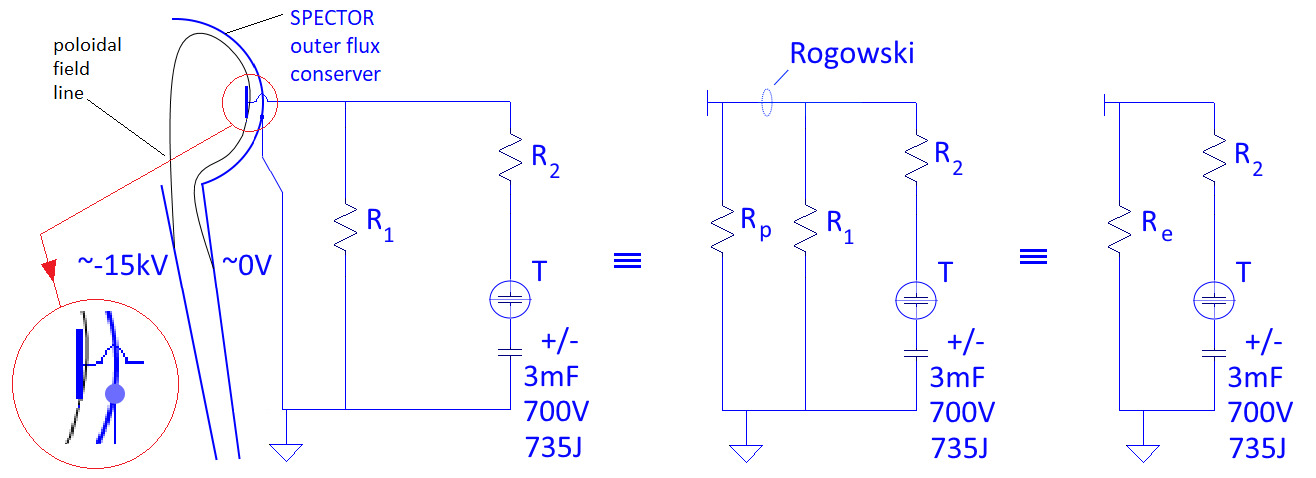}}}\textcolor{black}{\caption{\label{fig:Biasing-probe-circuit}$\,\,\,\,$Biasing probe circuit
schematic. Note the inset on the lower left of the figure depicts
the connection of the circuit to the flux conserver and probe.}
}
\end{figure}
\textcolor{black}{Figure \ref{fig:Biasing-probe-circuit} indicates
the most optimal of the biasing probe circuit configurations tested.
The biasing circuit was kept open-circuited until well after CT formation,
in order to protect biasing circuit components. A thyratron switch
(indicated in the figure) is robust against large amplitude negative
voltage spikes that can appear on the probe during CT formation when
initially open stuffing-field lines, that are resistively pinned to
the injector inner and outer electrodes, intersect the probe (see
left subfigure). These thyratron switches are designed to operate
at several kilovolts, and usually require several kiloamps of current
to remain closed, but careful setting of switch temperature enabled
operation at moderate voltages and currents. The biasing capacitor
voltage setting $V_{bc0}$, parallel and series resistors $R_{1}$
and $R_{2}$, and $R_{p}$, the plasma resistance between the electrode
($i.e.,$ probe) and flux conserver, determine $V_{probe}$, the voltage
measured between the probe and flux conserver, and $I_{probe},$ the
current drawn through the plasma edge. The current drawn has a radial
component, which leads, in the classical edge biasing scenario, to
$\mathbf{J}_{r}\times\mathbf{B}$ driven edge velocity shearing and
consequential decorrelation of turbulence cells and confinement improvement.
For the circuit with the $3\mbox{mF}$ capacitor depicted in figure
\ref{fig:Biasing-probe-circuit}, optimal circuit resistances were
found to be $R_{1}\sim0.2\Omega$, and $R_{2}\sim0.5\Omega$. Negative
electrode biasing was briefly tested; the results presented in this
paper were obtained with positive biasing. The effective resistance
$R_{e}$, comprised of $R_{p}$ and $R_{1}$ in parallel (see figure
\ref{fig:Biasing-probe-circuit}, right subfigure), is given by 
\begin{equation}
R_{e}(t)=\frac{R_{p}(t)\,R_{1}}{R_{p}(t)+R_{1}}\label{eq:1-1}
\end{equation}
The voltage applied by the capacitor on the probe is 
\begin{equation}
V_{applied}(t)=V_{bc}(t)\left(\frac{R_{e}(t)}{R_{e}(t)+R_{2}}\right)\label{eq:2}
\end{equation}
where $V_{bc}(t)$ is the voltage across the biasing capacitor. Equations
\ref{eq:1-1} and \ref{eq:2} can be combined to provide an expression
for $R_{p}$:
\begin{equation}
R_{p}(t)=\frac{R_{1}R_{2}V_{applied}(t)}{R_{1}(V_{bc}(t)-V_{applied}(t))-V_{applied}(t)\,R_{2}}\label{eq:3}
\end{equation}
}
\begin{figure}[H]
\centering{}\textcolor{black}{}\subfloat[]{\textcolor{black}{\includegraphics[width=8cm,height=5cm]{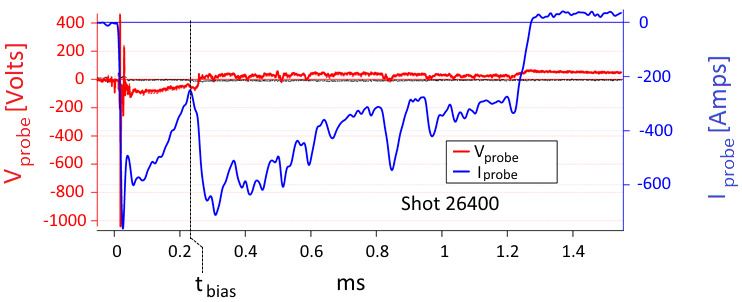}}}\textcolor{black}{\hfill{}}\subfloat[]{\textcolor{black}{\includegraphics[width=8cm,height=5cm]{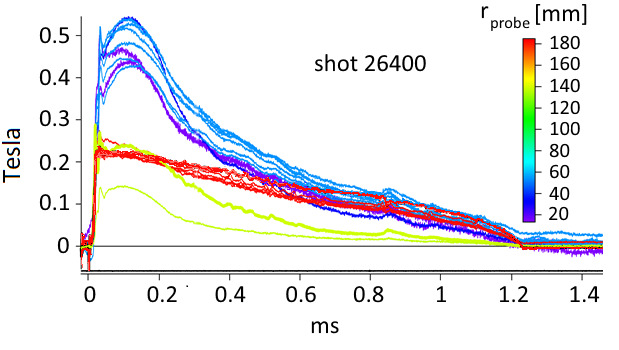}}}\textcolor{black}{\caption{\label{26400}$\,\,\,\,$Measured bias probe voltage and current (a),
and poloidal field (b) for shot 26400, which had $V_{bc0}=700$V (3mF
capacitor). Note that poloidal field data is colored by magnetic probe
radius.}
}
\end{figure}
\textcolor{black}{Figure \ref{26400}(a) shows the voltage measured
between the probe and the vacuum vessel, and the current drawn through
the plasma edge, as measured with the Rogowski coil indicated in figure
\ref{fig:Biasing-probe-circuit}, for shot 26400. At the biasing capacitor
voltage found to be most optimal for CT lifetime and electron temperature
(as obtained with the TS system), the voltage measured between the
probe and vacuum vessel was typically $V_{probe}\sim+50\mbox{V}$
to $+80\mbox{V}$, and the maximum radial current drawn to the probe
from the wall was $I_{probe}\sim700\mbox{A}$ to $\sim1\mbox{kA}$
shortly after firing the biasing capacitor(s). For shot 26400, the
electrode was inserted 11mm into the edge plasma, and biased at $t_{bias}=230\upmu\mbox{s}$
after firing the formation capacitor banks, as indicated in figure
\ref{26400}(a). Note that current is already flowing through the
plasma edge, and through resistor $R_{1},$ before $t_{bias}$, as
a result of the plasma-imposed potential on the electrode, which typically
led to a measurement of $V_{probe}\sim-100\mbox{V}$ when magnetized
plasma first enters the CT confinement area at around $20\upmu$s.
$V_{probe}$ and $I_{probe}$ decrease over time at a rate that depends
on plasma and circuit parameters. Figure \ref{26400}(b) indicates,
for shot 26400, the poloidal field measured at the magnetic probes
indicated as red dots in figure \ref{fig:Spector}(a). It is interesting
that the fluctuations in $B_{\theta}$, which are thought to be associated
with internal reconnection events, are also manifested on the biasing
voltage and current measurements, $e.g.,$ at $\sim845\upmu$s in
figures \ref{26400}(a) and (b). This observation is enabled by the
presence of the small parallel $R_{1}$. As edge plasma impedance
varies, as determined by internal MHD events, the system can divert
varying proportions of capacitor driven current through $R_{1}$.
In future studies, it may be possible to influence the behaviour of
the internal modes that cause the $B_{\theta}$ fluctuations, by driving
an edge current that is resonant with the fluctuations. }
\begin{figure}[H]
\begin{centering}
\textcolor{black}{}\subfloat[Biasing probe circuit diagram, including plasma voltage source ]{\textcolor{black}{\includegraphics[width=5.5cm,height=6.5cm]{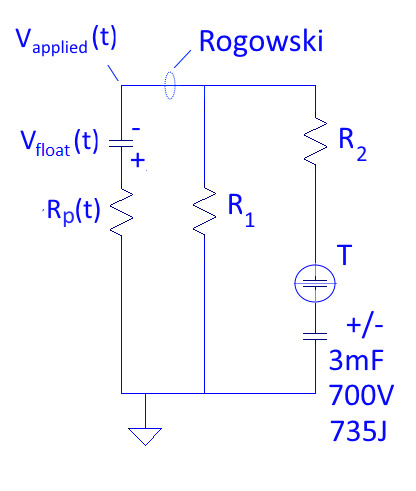}}}\textcolor{black}{\hfill{}}\subfloat[Current flow schematic for case with switch open]{\textcolor{black}{\includegraphics[width=4.4cm,height=7cm]{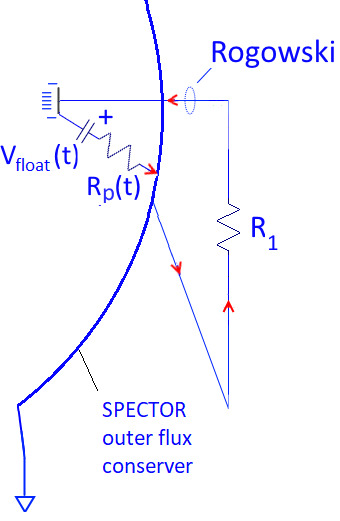}}}\textcolor{black}{\hfill{}}\subfloat[Current flow schematic for example case with switch closed and $R_{1}=2R_{p}$,
and $V_{bc0}>0$]{\textcolor{black}{\includegraphics[width=5.5cm,height=7cm]{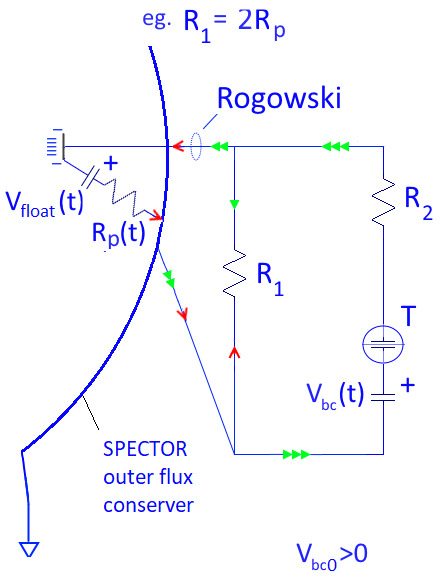}}}
\par\end{centering}
\centering{}\textcolor{black}{\caption{\label{fig:Circuit_plasma}$\,\,\,\,$Biasing probe circuit diagram,
including plasma voltage source, with current flow schematics }
}
\end{figure}
\textcolor{black}{When the plasma is considered as a time-dependent
voltage source, which biases the probe to floating potential $V_{float}(t)$,
a more complete circuit diagram is as depicted in figure \ref{fig:Circuit_plasma}(a).
The inclusion of $R_{1}$, a small external resistance in parallel
with $R_{p}$ (the plasma resistance between the probe and wall),
allows current driven by the floating potential to flow in the circuit
in the case where the thyratron switch is open (see figure \ref{fig:Circuit_plasma}(b)).
When the switch is closed, a proportion of the biasing capacitor-driven
current may divert to flow through $R_{1}$, see figure \ref{fig:Circuit_plasma}(c).
This proportion increases as $R_{p}$ increases with reducing electron
temperature as the CT decays, thereby allowing $I_{probe}$ to decrease
at a rate roughly in proportion to the rate of decrease of the main
CT plasma currents. The presence of an appropriately sized $R_{1}$
also prevents development of a sustained arc, which could damage the
wall and probe, through the ambient plasma that remains between the
probe and wall after the CT has extinguished. In previous edge biasing
studies on tokamaks, the standard is to maintain approximately constant
$V_{applied}$ and $I_{probe}$ for an extended time which is a segment
of the duration over which the approximately constant externally driven
toroidal plasma current flows. On SPECTOR plasmas, the plasma currents
are not driven and are allowed to decay naturally after formation,
so a circuit configuration that establishes constant $V_{applied}$
and $I_{probe}$ would not be compatible.}

\textcolor{black}{The differential voltage measured between the probe
and flux conserver is 
\begin{equation}
V_{probe}(t)=V_{applied}(t)+V_{float}(t)\label{eq:2.1}
\end{equation}
If the bias capacitor is not fired, and $R_{1}$ is removed from the
circuit, then in the open circuit condition $V_{probe}(t)=V_{float}(t)$.
Note that $V_{float}$ is not measured directly on each shot, however,
looking at the $V_{probe}$ measurements taken during several open
circuit, probe-in shots, the floating potential can be approximated
as an RC rise of the form 
\begin{equation}
V_{float}(t)=V_{f0}\,e^{-\frac{t}{\tau_{RCf}}}\label{eq:2.11}
\end{equation}
with $V_{f0}\sim-80\mbox{V}$, and, (depending on CT lifetime) $\tau_{RCf}\sim1\mbox{ms}$.
$V_{float}(t)$ rises from $\sim-80\mbox{V}$ at the time when plasma
enters the CT confinement region, to $0\mbox{V}$ when the CT has
decayed away. With this, an approximation for $V_{applied}$ can be
made using equation \ref{eq:2.1}. $V_{bc}(t)$, the voltage across
the biasing capacitor, was not measured directly in the experiment,
but can be estimated as 
\begin{equation}
V_{bc}(t)=V_{bc0}\,e^{-\frac{t}{\tau_{RCb}}}\label{eq:2.12}
\end{equation}
 where, for shot 26400, $V_{bc0}=700\mbox{V}$ and $\tau_{RCb}=0.5\Omega*3\mbox{mF}=1.5\mbox{ms}$
(resistance $R_{2}=0.5\Omega\gg R_{e})$. With these approximations
for $V_{applied}(t)$ and $V_{bc}(t)$, an estimate of $R_{p}(t)$,
the plasma resistance between the probe and flux conserver, is evaluated
using equation \ref{eq:3}, between $t_{bias}=250\upmu\mbox{s}$ until
the time when the CT has decayed.  }
\begin{figure}[H]
\centering{}\textcolor{black}{}\subfloat[]{\textcolor{black}{\includegraphics[width=5.3cm,height=5cm]{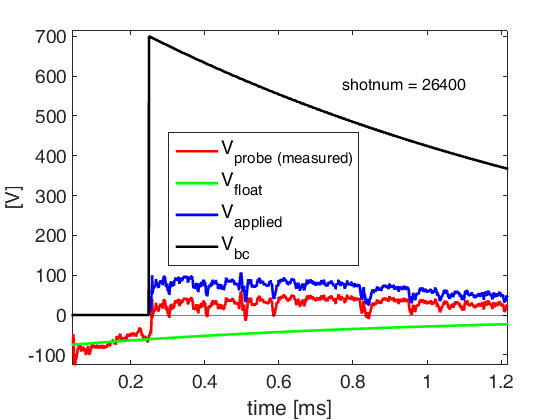}}}\textcolor{black}{\hfill{}}\subfloat[]{\textcolor{black}{\includegraphics[width=4.7cm,height=5cm]{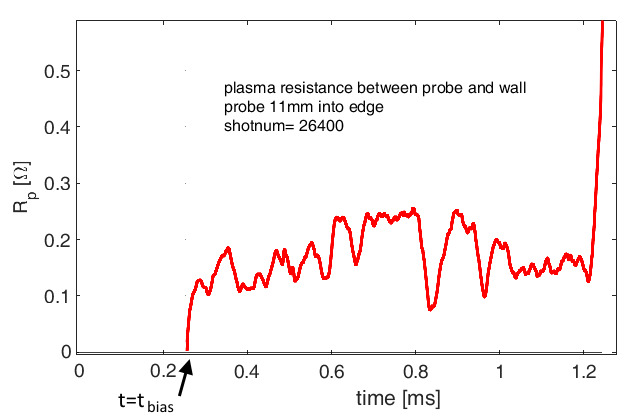}}}\textcolor{black}{\hfill{}}\subfloat[]{\textcolor{black}{\includegraphics[width=5.5cm,height=5cm]{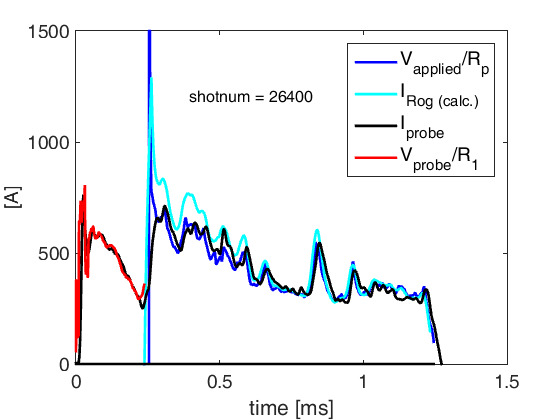}}}\textcolor{black}{\caption{\label{26400VRI_bias}$\,\,\,\,$Calculated / measured probe voltages,
edge plasma resistance and bias currents for shot 26400}
}
\end{figure}
\textcolor{black}{The approximations (from equations \ref{eq:2.1},
\ref{eq:2.11}, and \ref{eq:2.12}) for $V_{applied}(t)$, $V_{float}(t)$,
and $V_{bc}(t)$, and measured $V_{probe}(t)$, for shot 26400, are
shown in figure \ref{26400VRI_bias}(a). Figure \ref{26400VRI_bias}(b)
shows the estimation, from equation \ref{eq:3}, for $R_{p}(t)$.
$R_{p}(t)\sim0.15\Omega$ to $0.2\Omega$, and rises as $T_{e}$ decreases
($\eta_{plasma}$ increases) over CT decay, then drops as the edge
current path length $L$ (recall $R(t)=\eta(t)L(t)/A(t)$) decreases.
Note that the current path will have a principal component along the
helical magnetic field between the probe (with insertion depth 11mm)
and the flux conserver. Path length decreases because $B_{\theta}$
decreases faster than $B_{\phi}$ (CT toroidal field is maintained
at a relatively constant level by the crow-barred external shaft current)
as the CT decays, $i.e.,$ $q$ increases - there are fewer poloidal
transits for each toroidal transit along the path which defines $R_{p}$.
The sharp dip in $R_{p}$ at $t\sim845\upmu\mbox{s}$ coincides with
the fluctuations in $I_{probe}(t)$ and $B_{\theta}$ seen in figures
\ref{26400}(a) and (b). Note that the current through the path enclosed
by the Rogowski coil depicted in figures \ref{fig:Biasing-probe-circuit}
and \ref{fig:Circuit_plasma} can be calculated using basic circuit
theory as: 
\begin{equation}
I_{rog\,(calc.)}=\frac{1}{R_{p}(t)}\left[\frac{R_{2}\left(V_{bc}(t)\,R_{1}+V_{float}(t)\,R_{1}+V_{bc}(t)\,R_{p}(t)\right)}{R_{1}R_{2}+R_{2}R_{p}(t)+R_{p}(t)\,R_{1}}-V_{bc}(t)-V_{float}(t)\right]\label{eq:4}
\end{equation}
Figure \ref{26400VRI_bias}(c) compares measured $I_{probe}(t)$ (black
trace) with calculated parameters, to verify the calculation of $R_{p}(t)$.
Referring to figure \ref{fig:Circuit_plasma}(c), it is seen that
$V_{applied}(t)/R_{p}(t)$ should, as is confirmed in figure \ref{26400VRI_bias}(c)
(dark blue trace), give the measured $I_{probe}(t)$ current when
the switch is closed after $t=t_{bias}$. Referring to figure \ref{fig:Circuit_plasma}(b),
$V_{probe}(t)/R_{1}\sim I_{probe}(t)$ when the switch is open before
$t=t_{bias}$ (red trace in \ref{26400VRI_bias}(c)). A good match
to measured $I_{probe}(t)$ is found by using calculated $R_{p}(t)$
and the estimated profile of $V_{float}(t)$ in equation \ref{eq:4}
(after $t=t_{bias}$, cyan trace). A good estimate of $R_{p}(t)$
is useful for optimizing external circuit resistances.}

\section{\textcolor{black}{Main results\label{sec:Main-results} }}

\textcolor{black}{}
\begin{figure}[H]
\centering{}\textcolor{black}{}\subfloat[]{\textcolor{black}{\includegraphics[width=8cm,height=6cm]{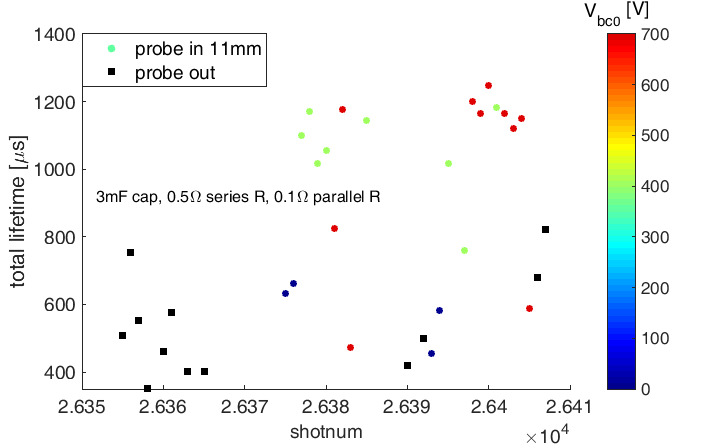}}}\textcolor{black}{\hfill{}}\subfloat[]{\textcolor{black}{\includegraphics[width=8cm,height=6cm]{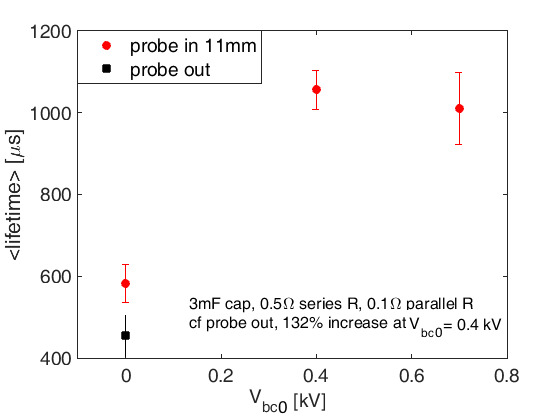}}
\raggedleft{}\textcolor{black}{}}\textcolor{black}{\caption{\label{fig:life_3mF_se}$\,\,\,\,$$3\mbox{mF}$ capacitor: total
CT lifetimes $cf.\,\,V_{bc0}$ (bias capacitor voltage setting)}
}
\end{figure}
\textcolor{black}{Figure \ref{fig:life_3mF_se}(a) indicates how CT
lifetimes varied with $V_{bc0}$ (coloured circles) for shots taken
with the biasing probe inserted $11\mbox{mm}$ into the plasma edge
in the configuration using the $3\mbox{mF}$ biasing capacitor circuit,
with $R_{1}=0.1\Omega$ and $R_{2}=0.5\Omega$, compared with shots
taken with the probe removed (black squares). Figure \ref{fig:life_3mF_se}(b)
indicates the average of CT lifetimes for the probe-out configuration
(black squares), and the averages for the probe-in configuration (red
circles) for the setpoints $V_{bc0}=0\mbox{V},\,400$V, and $700$V,
where the error bars are defined with the standard error.  It is indicated
that CT lifetime increased from around 450 to 600eV even when the
biasing capacitor was not fired - in that case, the presence of the
resistor ($R_{1}$) in parallel with the biasing capacitor enables
current, driven by the potential applied by the plasma, to flow from
the electrode to the wall. At $V_{bc0}=400$V, CT lifetime increased
by a factor of around 2.3, from $\sim460\upmu$s to $\sim1070\upmu$s.
Note that TS data is not available for the configuration with the
3mF capacitor in the biasing circuit.}
\begin{figure}[H]
\centering{}\textcolor{black}{}\subfloat[]{\textcolor{black}{\includegraphics[width=8cm,height=6cm]{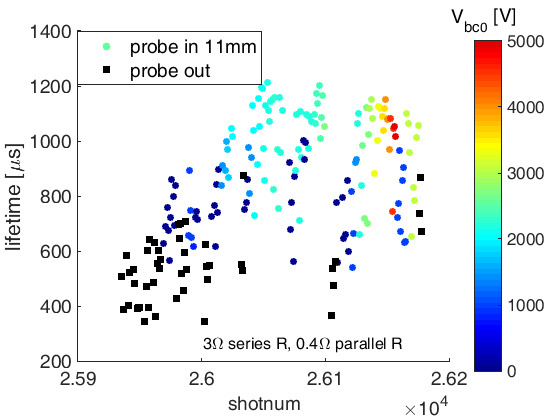}}}\textcolor{black}{\hfill{}}\subfloat[]{\textcolor{black}{\includegraphics[width=8cm,height=6cm]{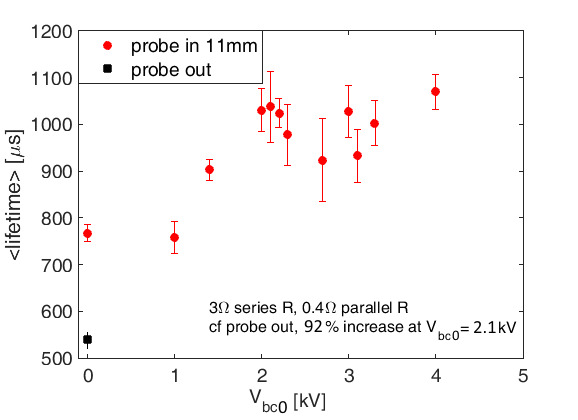}}
\raggedleft{}\textcolor{black}{}}\textcolor{black}{\caption{\label{fig:life_100uF_se}$\,\,\,\,$$100\upmu$F capacitor: CT lifetimes
$cf.\,\,V_{bc0}$ }
}
\end{figure}
\textcolor{black}{Figure \ref{fig:life_100uF_se} shows equivalent
information for shots taken with a $100\upmu$F, 5kV capacitor in
the biasing circuit, with $R_{1}=0.4\Omega$ and $R_{2}=3\Omega$
(TS data is available for this configuration). CT lifetimes were approximately
doubled in this configuration, with an optimal biasing capacitor setpoint
of $V_{bc0}\sim2$kV. }
\begin{figure}[H]
\centering{}\textcolor{black}{}\subfloat[]{\textcolor{black}{\includegraphics[width=8cm,height=6cm]{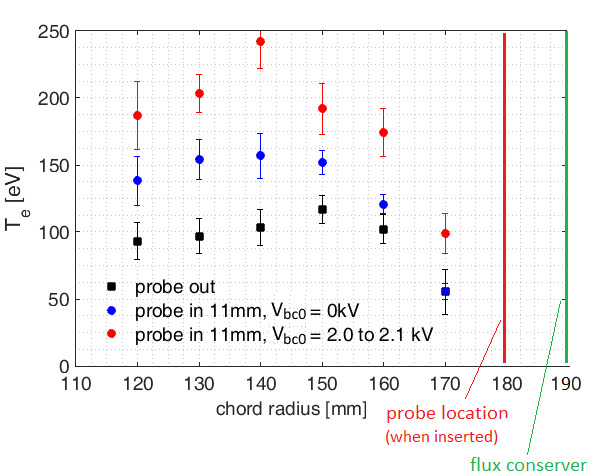}}}\textcolor{black}{\hfill{}}\subfloat[]{\textcolor{black}{\includegraphics[width=8cm,height=6cm]{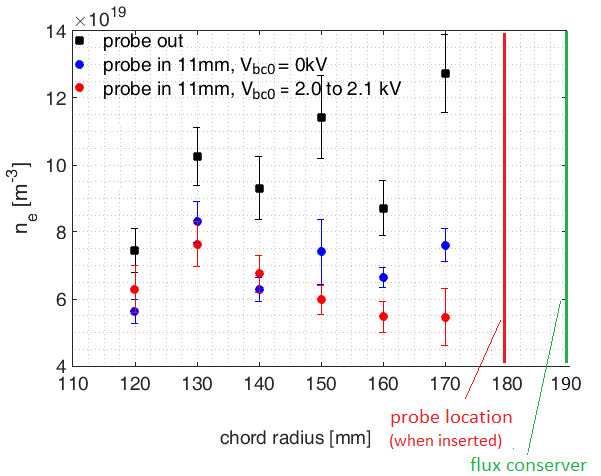}}
\raggedleft{}\textcolor{black}{}}\textcolor{black}{\caption{\label{fig:TS_2profile}$\,\,\,\,$$100\upmu$F capacitor: electron
temperature and density profiles from the Thomson scattering diagnostic
at $300\upmu$s, for $V_{bc0}\sim2$$\mbox{kV}$. Note that the Thomson
scattering sampling points are depicted in figure \ref{fig: Spect_n_Te}(b).}
}
\end{figure}
\textcolor{black}{Figure \ref{fig:TS_2profile} shows shot data indicating
the temperature and density profiles obtained with the TS system at
$300\upmu$s after firing the formation capacitor banks, for the configuration
with the $100\upmu$F, 5kV biasing capacitor. Note that the TS sampling
points are indicated in figure \ref{fig: Spect_n_Te}(b). With $V_{bc0}\sim2$kV,
the measurements indicate that temperature is more than doubled at
the inner sampling points, increasing by a factor of around 2.4 at
the sampling point at $r=140$mm, (black squares $cf.$ red circles)
and the proportional increase in temperature falls off towards the
CT edge. Note that current drawn through the CT edge leads to a temperature
increase even when no voltage is externally applied to the electrode
(black squares $cf.$ blue circles). Referring to figure \ref{fig:TS_2profile}(b),
electron density is markedly reduced when the electrode is inserted
and the reduction is enhanced when the electrode is externally biased.
It can be seen how the density profile is quite hollow without biasing,
with the density highest towards the plasma edge. With biasing, a
centrally peaked density profile is observed. Without biasing, the
electron temperature profile is peaked at around 150mm (figure \ref{fig:TS_2profile}(a)).
The electron temperature profile become more peaked, and the peak
moves inwards to around 140mm when biasing is applied. It may be inferred
that the pressure profile also becomes more centrally-peaked with
biasing. The proportional decrease in density is greater towards the
CT edge, consistent with the theory that edge fueling impediment due
to an edge transport barrier is largely responsible for the density
reduction (see appendix \ref{sec:Neutral_SPECTOR}). The diagnostic
indicates an electron density reduction by factors of approximately
1.5 and 2.3 at $r=130$mm and $r=170$mm respectively. }
\begin{figure}[H]
\centering{}\textcolor{black}{}\subfloat[]{\textcolor{black}{\includegraphics[width=8cm,height=5cm]{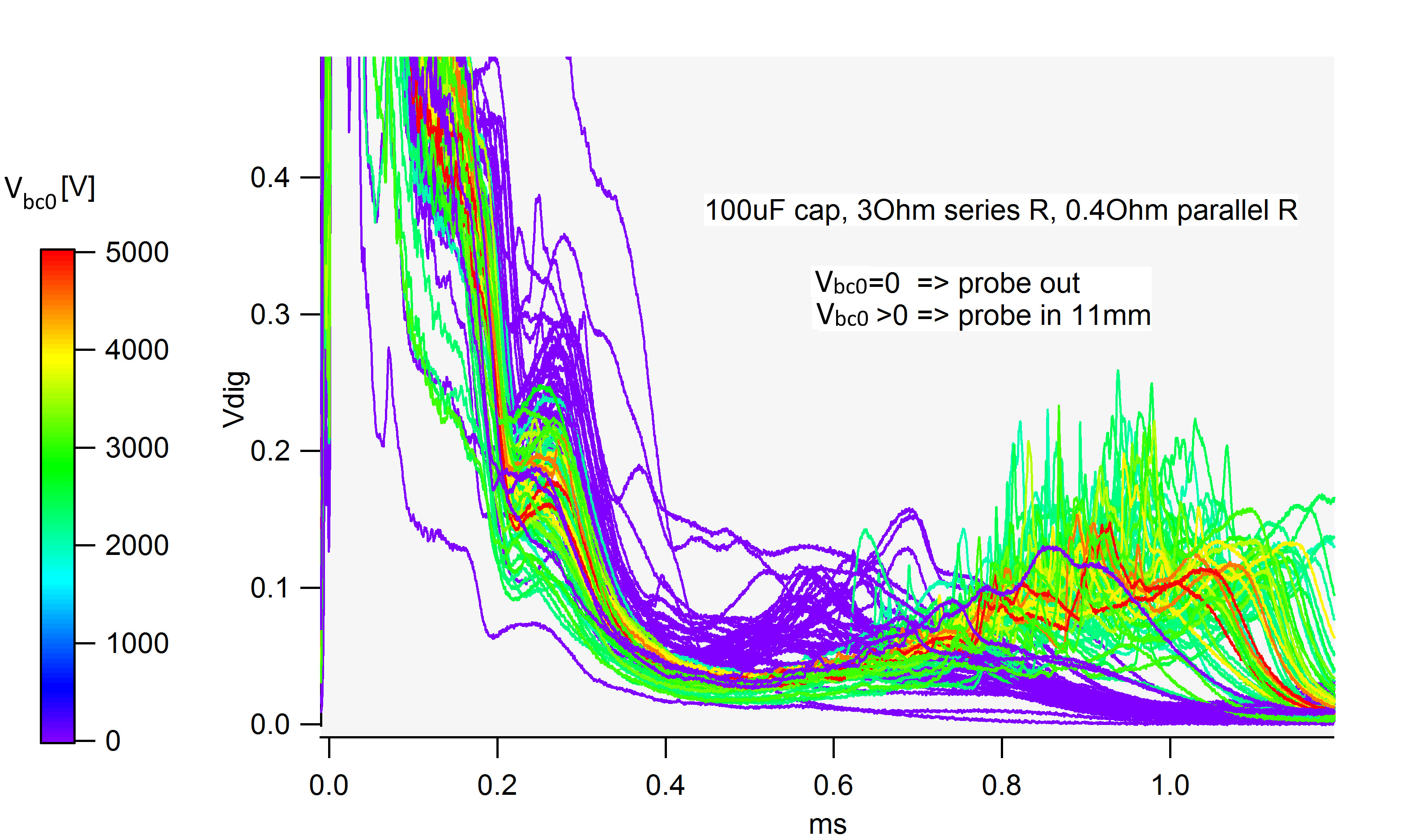}}}\textcolor{black}{\hfill{}}\subfloat[]{\textcolor{black}{\includegraphics[width=8cm,height=5cm]{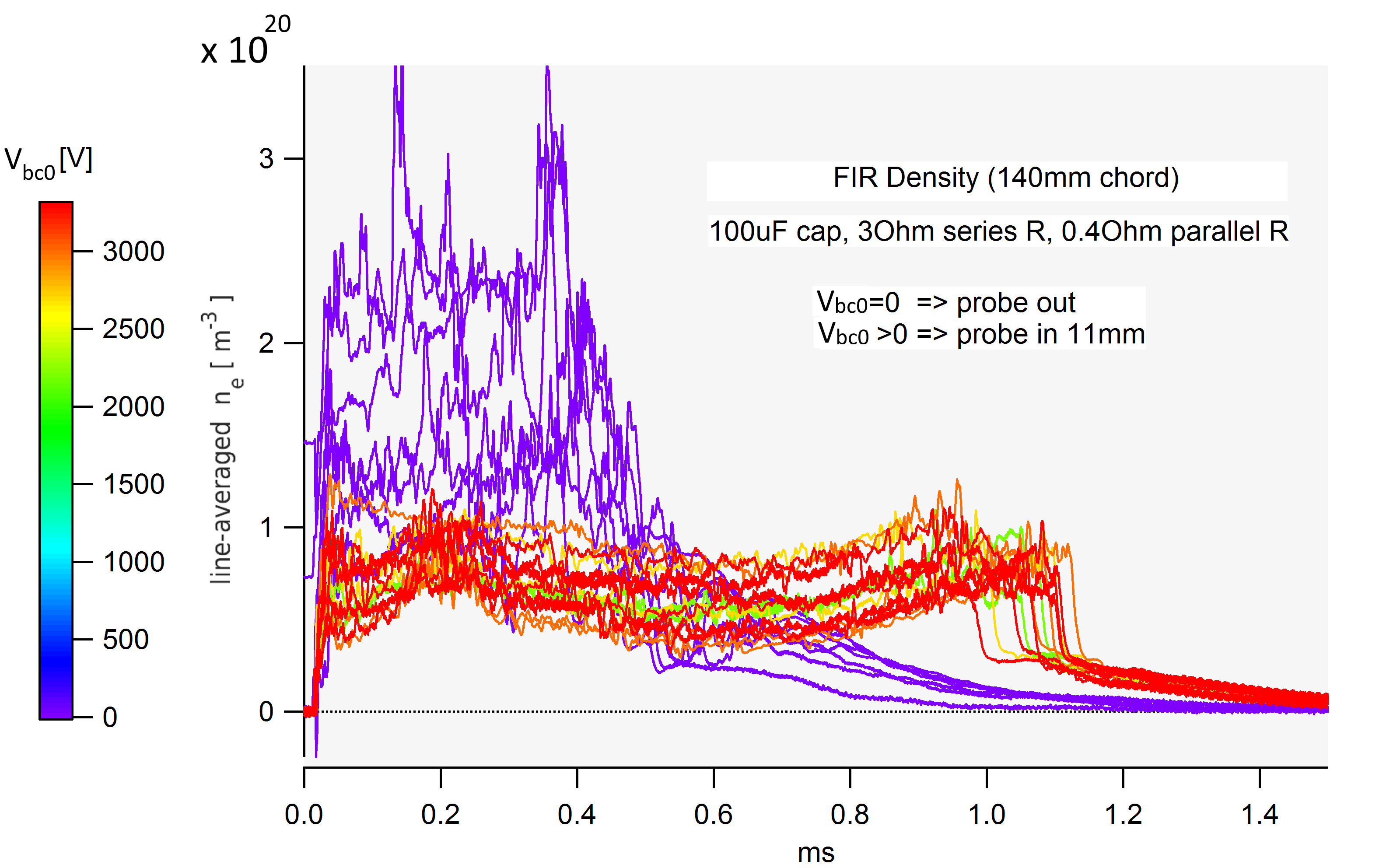}}
\raggedleft{}\textcolor{black}{}}\textcolor{black}{\caption{\label{fig:Halphavert}$\,\,\,\,$$100\upmu$F capacitor: (a) $\mbox{H}_{\alpha}$
intensity, (b) electron density (FIR interferometer)}
}
\end{figure}
\textcolor{black}{Figure \ref{fig:Halphavert}(a) indicates how $\mbox{H}_{\alpha}$
intensity, along a vertical chord located at $r=88$mm, is reduced
when the electrode is inserted and biased. Ions and electrons, that
diffuse out of the plasma to the vessel walls, cool and recombine
at the walls to form neutral atoms, which can then exhibit $\mbox{H}_{\alpha}$
emission, before being re-ionized and $\mbox{recycled}$ back into
the plasma as cold ions and electrons. The presence of a transport
barrier reduces the level of $\mbox{H}_{\alpha}$ emission by reducing
the flux of charged particles to the vessel walls. $\mbox{H}_{\alpha}$
intensity reduction is a sign of reduced recombination at the vessel
walls \cite{Taylor_CCT}. It is thought that the reduction of the
level of recombination observed may be associated with a establishment
of a transport barrier due to shearing effects, and consequent reduction
of flux of plasma particles from the CT core towards the vessel walls.
 The purple traces are from shots with the electrode removed from
the vacuum vessel. As shown in figure \ref{fig:Halphavert}(b), time-resolved
FIR interferometer data from the chord at 140mm (FIR chord locations
are indicated in figure \ref{fig: Spect_n_Te}(a)) confirms the reduction
in electron density when the biased probe is inserted into the plasma
edge. Again, the purple traces are from shots taken with the electrode
retracted. This density reduction is thought to be due to the effect
of the transport barrier impeding the level of CT fueling associated
with neutral gas diffusing up the gun, as discussed in appendix \ref{sec:Neutral_SPECTOR}.
Note that the fueling effect is not entirely eliminated by the biasing
effect - density starts to increase at around $500$ to 600$\upmu$s
($cf.$ figure \ref{fig:Spector_nN}(b)). Note that H$_{\alpha}$
intensity increases dramatically at around the same time (figure \ref{fig:Halphavert}(a)).
The CTs associated with the purple traces (probe-out configuration)
in figure \ref{fig:Halphavert}(a) and (b) do not last for long enough
to enable observation of the density and H$_{\alpha}$ intensity increases
at that time. }

\section{\textcolor{black}{Discussion and conclusions\label{sec:Discussion-and-conclusions}}}

\textcolor{black}{Significant increases in CT lifetime and electron
temperature, and reductions in electron density and $\mbox{H}_{\alpha}$
intensity, were observed when the electrode was inserted into the
plasma edge, even when the biasing capacitor was not fired. In that
case, the presence of the resistor ($R_{1}$) in parallel with the
biasing capacitor enables current, driven by the potential applied
by the plasma, to flow from the electrode to the wall. Note that in
cases where the biasing capacitor was not fired, the enhanced performance
was eliminated when $R_{1}$ was removed from the circuit. In terms
of enhanced CT lifetime, which was observed to increase by a factor
of up to 2.3, the optimal biasing circuit tested was with the 3mF
capacitor in place, but TS data was not available in that configuration.
Up to $\sim1200$A was drawn $11\mbox{\mbox{mm}}$ through the edge
plasma, while improving CT lifetime and temperature. CT lifetimes
and electron temperatures were observed to increase by factors of
around 2 and 2.4 (temperature near the CT core) respectively in the
configuration with the 100$\upmu$F capacitor charged to 2.1kV, while
density decreased by a factor of around 2.3 near the CT edge. This
density reduction and peaking of the originally hollow density profile
is thought to be due to the effect of a transport barrier, induced
by the effect of the velocity shear caused by the imposed radial electric
field, impeding the level of CT fueling associated with neutral gas
diffusing up the gun (see appendix \ref{sec:Neutral_SPECTOR}). The
consequent reduction of cool particle flux towards the CT interior
is thought to be largely responsible for the particularly significant
increases in observed temperature. In contrast, prior edge biasing
experiments in tokamaks generally show an increase in density with
no significant change in temperature. In those cases, the transport
barrier results in improved energy and particle confinement, but the
increase in energy is offset by the increase in particle count, so
that convincing temperature increases are not observed. }

\textcolor{black}{The total power associated with the biasing can
be estimated as $P_{bias}=V_{probe}I_{probe}$ . For shot 26400 (figure
\ref{26400}(a)), this can be estimated as $P_{bias}\sim35$kW (at
300$\upmu$s, when $V_{probe}\sim50$V and $I_{probe}\sim700$A. Typical
loop voltages associated with the SPECTOR injector are of the order
of 3 Volts, while plasma current was around 300kA for shot 26400 at
300$\upmu$s, so that the Ohmic heating power is approximately $P_{\Omega}\sim$1MW.
It appears that any plasma heating associated with the power injected
by biasing would not be significant compared with the Ohmic heating
power, and cannot be responsible for the observed increases in electron
temperature. }

\textcolor{black}{Note that the biasing experiment was conducted without
a fresh lithium coating on the inside of the SPECTOR flux conserver.
With a fresh coating, CT lifetimes are typically around 2ms. The biasing
experiment may be run again with a fresh coating. The experiment was
conducted over a short period (less than two weeks). As the majority
of the probe-out shots were taken at the beginning of each day, there
is likely some data skew due to cleaning effects. The improvement
shown with biasing may be extended with further circuit optimization.
Negative biasing was tested briefly - a slight increase of electron
temperature and a peaking of the electron temperature profile was
observed, but there was no evidence of lifetime increase. It may be
that the ion-sputtering of the probe associated with negative biasing
lead to performance degradation associated with plasma impurities
that offset the improvement associated with the establishment of a
transport barrier. Perhaps more cleaning shots are required to see
a significant improvement with negative biasing - the efficacy of
negative biasing hasn't been confirmed. An IV curve was p=\nobreakdash-=\nobreakdash-roduced
with the electrode biased to a range of positive and negative voltages
on a shot to shot basis. Langmuir analysis indicated $T_{e}\sim130$$\mbox{\mbox{eV}}$
and $n_{e}\sim10^{19}[\mbox{m}^{-3}]$ at the probe location at $300$$\upmu\mbox{s}$,
and $T_{e}\sim85$$\mbox{\mbox{eV}}$ and $n_{e}\sim5\times10^{18}[\mbox{m}^{-3}]$
at $600$$\upmu\mbox{s}$. Compared with TS data, the electron temperature
estimates in particular appear too high. The fact that probe biasing
affects electron temperature and electron density makes the Langmuir
analysis results dubious at best, but it may be possible to correct
for this effect. }

\textcolor{black}{It would be worth repeating the experiment with
a fresh lithium coating on the inner flux conserver. Circuit parameters,
probe insertion depth, and machine operation settings should be optimized
further. Future SPECTOR biasing experiments would ideally have more
diagnostics available so that the effects of biasing on edge conditions
(for example, profiles of electric field, density, temperature edge,
and plasma velocity) can be characterised using Langmuir and Mach
probes, and ion Doppler diagnostics. Negative biasing may be tested
more rigorously. It would be interesting to look at the effects of
driving edge current resonant to the MHD behaviour that manifests
itself in the form of fluctuations on measurements including CT poloidal
field.}

\textcolor{black}{The biasing experiment was especially noteworthy
because it has generally been found that insertion of foreign objects,
such as thin alumina tubes containing magnetic probes, into SPECTOR
CTs, leads to performance degradation associated with plasma impurities.
After the extensive problems encountered relating to plasma/material
interaction and impurities during the magnetic compression experiment
\cite{thesis,exppaper}, special care was taken to choose a plasma-compatible
material for the biasing electrode assembly. The pyrolytic boron nitride
tube and molybdenum electrode combination seems to have been a good
choice - at least the benefit due to drawing a current through the
CT edge outweighed any performance degradation that may have been
associated with impurities introduced to the system.}

\section{\textcolor{black}{Acknowledgments}}

\textcolor{black}{Funding was provided in part by General Fusion Inc.,
Mitacs, University of Saskatchewan, and NSERC. Particular thanks to
Alex Mossman, Kelly Epp, Akbar Rohollahi, Russ Ivanov, and Adrian
Wong for machine operation support, to Russ Ivanov, Ivan Khalzov,
and Meritt Reynolds for useful discussions, and to Wade Zawalski,
Curtis Gutjahr, Blake Rablah, James Wilkie, Alan Read, Mark Bunce,
Pat Carle and Bill Young for hardware and diagnostics support.}

\appendix

\section{\textcolor{black}{Appendix: Simulation of interaction between neutral
and plasma fluids in SPECTOR geometry\label{sec:Neutral_SPECTOR} }}

\textcolor{black}{It is usual to observe a significant rise in electron
density at around $500\upmu$s on the SPECTOR machine, and it is thought
that this may be a result of neutral gas, that remains around the
gas valve locations after CT formation, diffusing up the gun. Ionization
of the neutral particles would lead to CT fueling and an increase
in observed electron density. An energy, particle, and toroidal flux
conserving finite element axisymmetric MHD code was developed to study
CT formation into a levitation field, and magnetic compression \cite{thesis,exppaper,SIMpaper}.
The Braginskii MHD equations with anisotropic heat conduction were
implemented. As described in \cite{thesis,Neut_paper}, a plasma-neutral
interaction model including ionization, recombination, charge-exchange
reactions, and a neutral particle source, was implemented to the MHD
code and used to study the effect of neutral gas on simulated CT formation
in SPECTOR geometry. }

\textcolor{black}{}
\begin{figure}[H]
\textcolor{black}{}\subfloat[]{\raggedright{}\textcolor{black}{\includegraphics[scale=0.5]{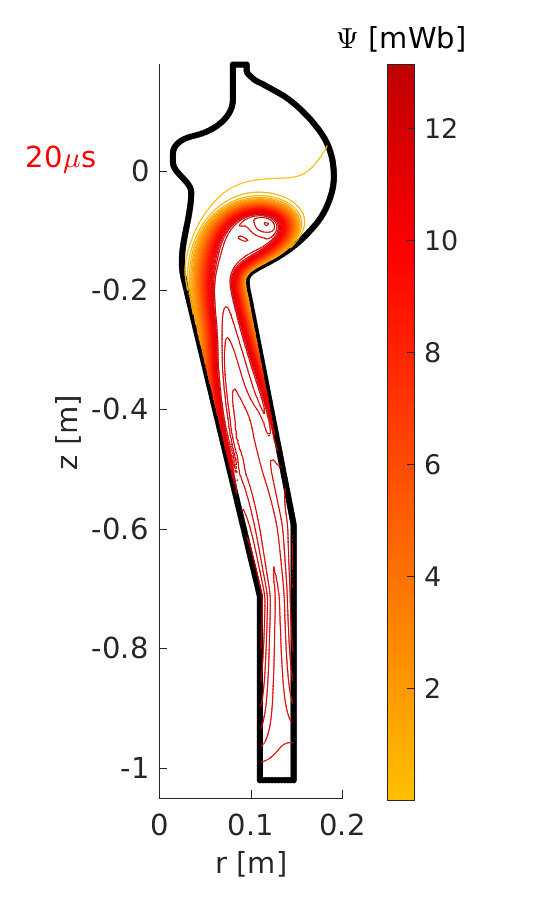}}}\textcolor{black}{\hfill{}}\subfloat[]{\raggedright{}\textcolor{black}{\includegraphics[scale=0.5]{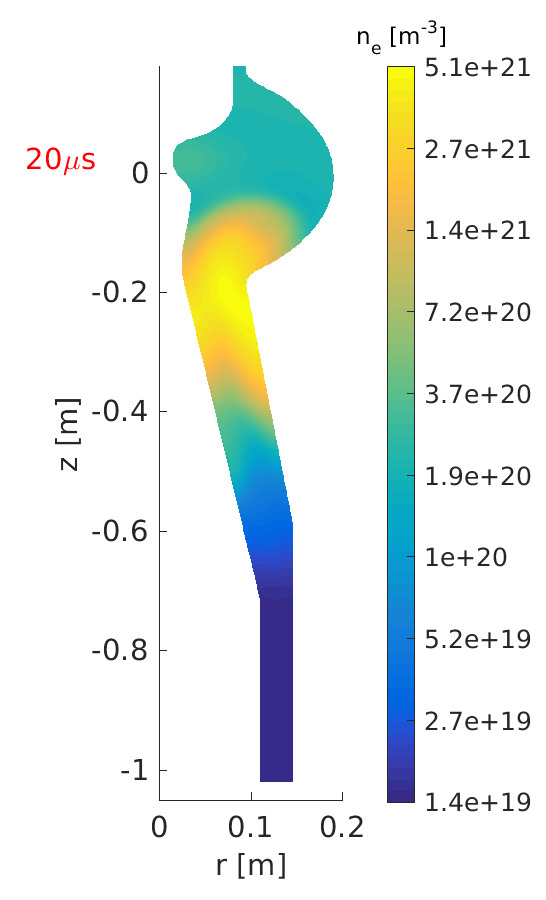}}}\textcolor{black}{\hfill{}}\subfloat[]{\raggedright{}\textcolor{black}{\includegraphics[scale=0.5]{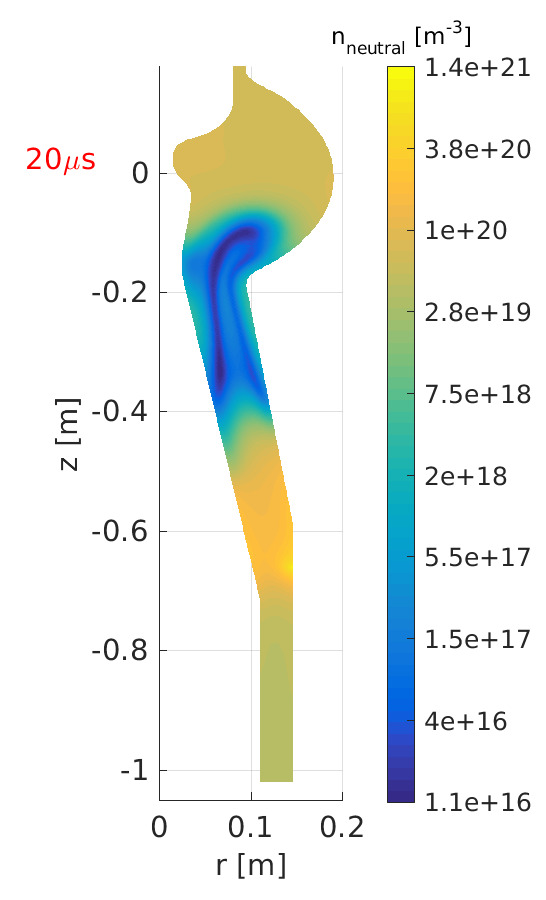}}}

\textcolor{black}{}\subfloat[]{\raggedright{}\textcolor{black}{\includegraphics[scale=0.5]{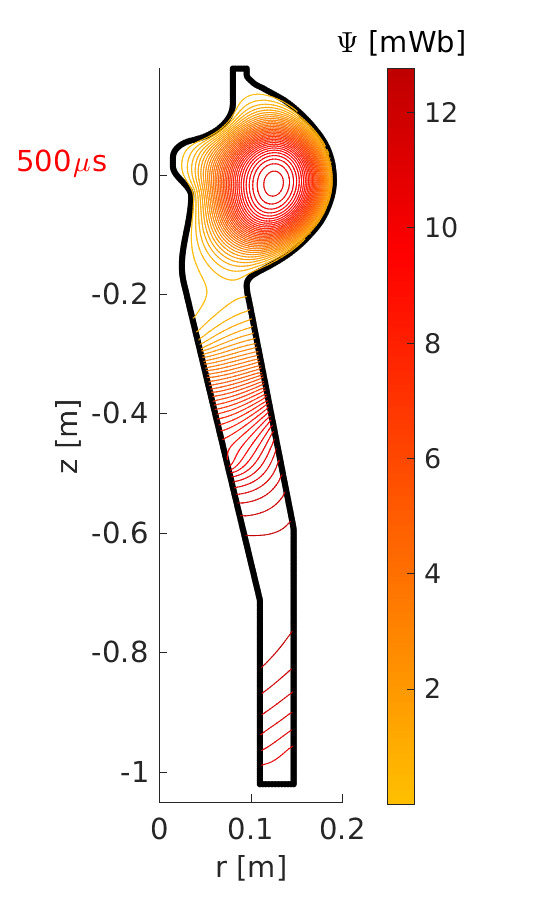}}}\textcolor{black}{\hfill{}}\subfloat[]{\raggedright{}\textcolor{black}{\includegraphics[scale=0.5]{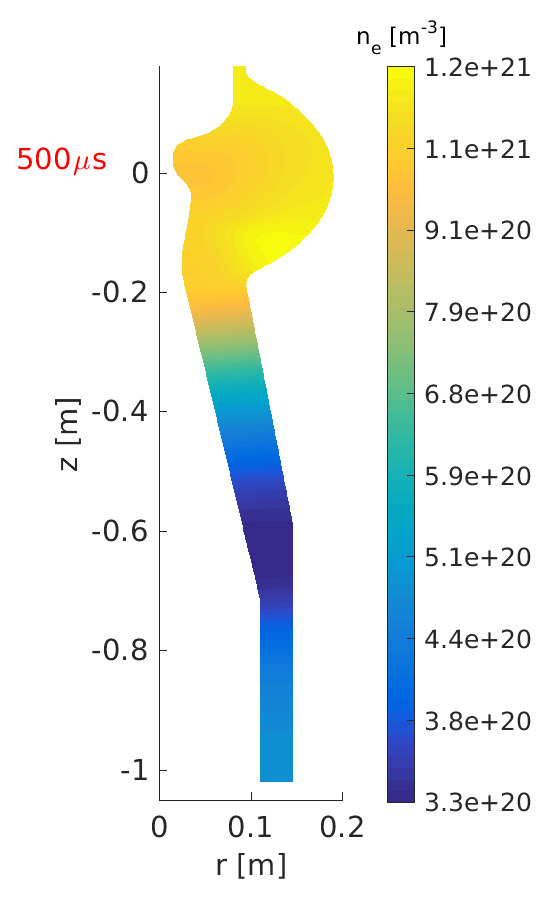}}}\textcolor{black}{\hfill{}}\subfloat[]{\raggedright{}\textcolor{black}{\includegraphics[scale=0.5]{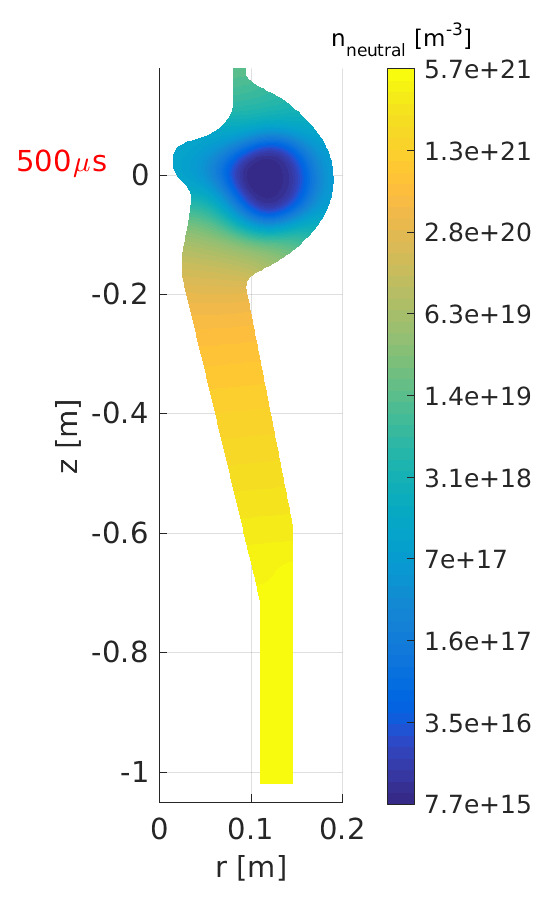}}}

\textcolor{black}{\caption{\label{fig:SPEC_psi_n_nN}$\,\,\,\,$Poloidal flux contours and profiles
of electron and neutral fluid densities at various times from a simulation
of CT formation in the SPECTOR plasma injector}
}
\end{figure}
\textcolor{black}{Figures \ref{fig:SPEC_psi_n_nN}(a), (b) and (c)
show $\psi$ contours and profiles of $n_{e}$ and $n_{n}$ at $20\upmu$s,
as plasma enters the CT containment region. Profiles of the same quantities
are shown in figures \ref{fig:SPEC_psi_n_nN}(d), (e) and (f) at $500\upmu$s,
around the time when the rise in measured electron density is usually
observed. It can be seen how neutral fluid density is highest at the
bottom of the gun barrel - any neutral gas advected or diffusing upwards
is ionized. A region of particularly high electron density is apparent
just above, and outboard of, the entrance to the containment region
- this is due to the fueling effect arising from neutral gas diffusion.
}\\
\textcolor{black}{}
\begin{figure}[H]
\textcolor{black}{}\subfloat[]{\raggedright{}\textcolor{black}{\includegraphics[width=7cm,height=6.3cm]{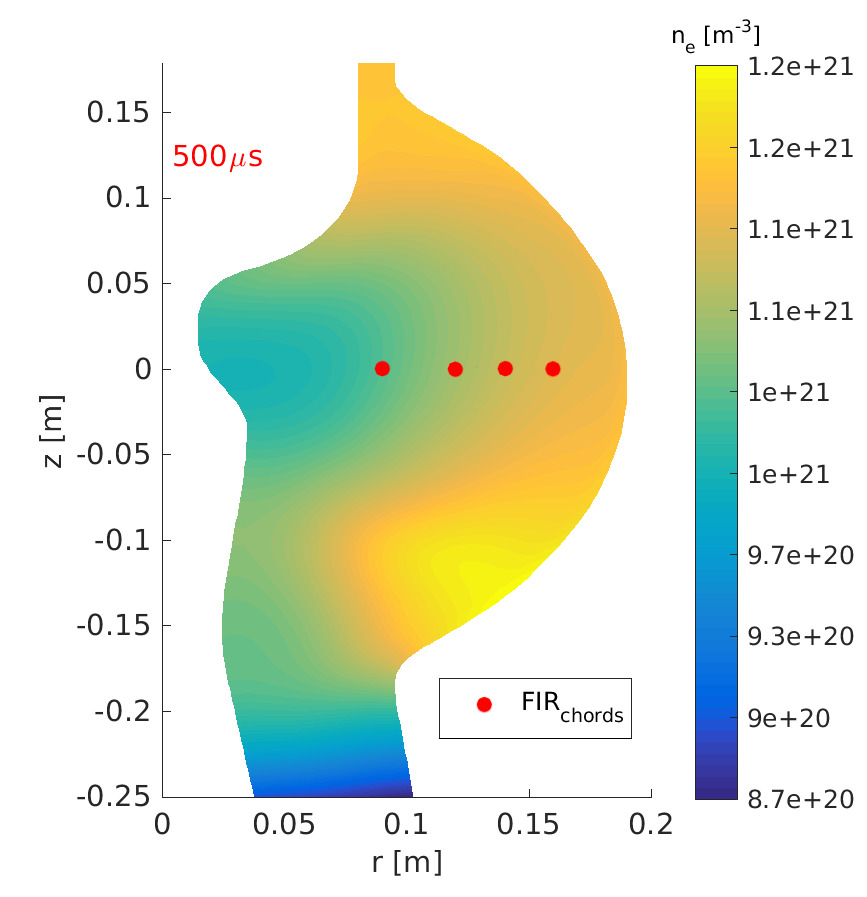}}}\textcolor{black}{\hfill{}}\subfloat[]{\raggedright{}\textcolor{black}{\includegraphics[width=7cm,height=6.3cm]{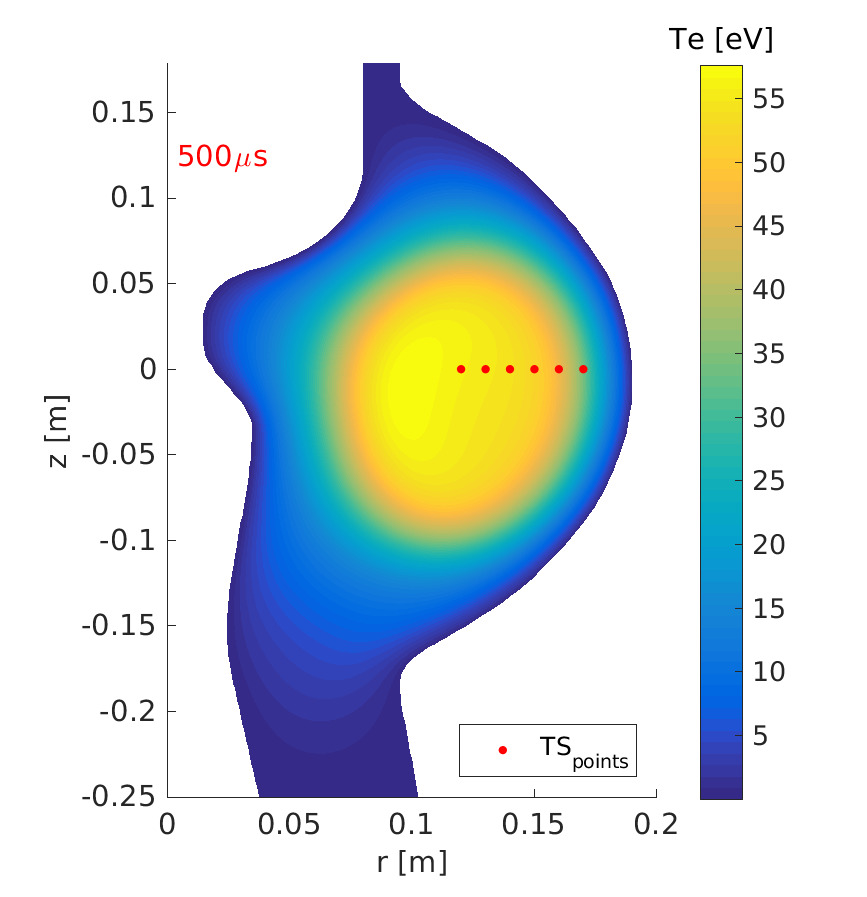}}}

\textcolor{black}{\caption{\label{fig: Spect_n_Te}$\,\,\,\,$Profiles of electron density and
temperature at $500\upmu$s from a simulation of CT formation in the
SPECTOR plasma injector}
}
\end{figure}
\textcolor{black}{The region of particularly high electron density
is more defined in figure \ref{fig: Spect_n_Te}(a), in which cross-sections
of the horizontal chords representing the lines of sight of the FIR
(far-infrared) interferometer \cite{polarimetryGF} are also depicted.
The electron temperature profile at 500$\upmu$s is shown in figure
\ref{fig: Spect_n_Te}(b). Referring to figure \ref{fig:SPEC_psi_n_nN}(f),
it can be seen how neutral fluid density is low in regions of high
$T_{e}$ as a result of ionization.}

\textcolor{black}{}
\begin{figure}[H]
\textcolor{black}{}\subfloat[]{\raggedright{}\textcolor{black}{\includegraphics[width=7cm,height=5cm]{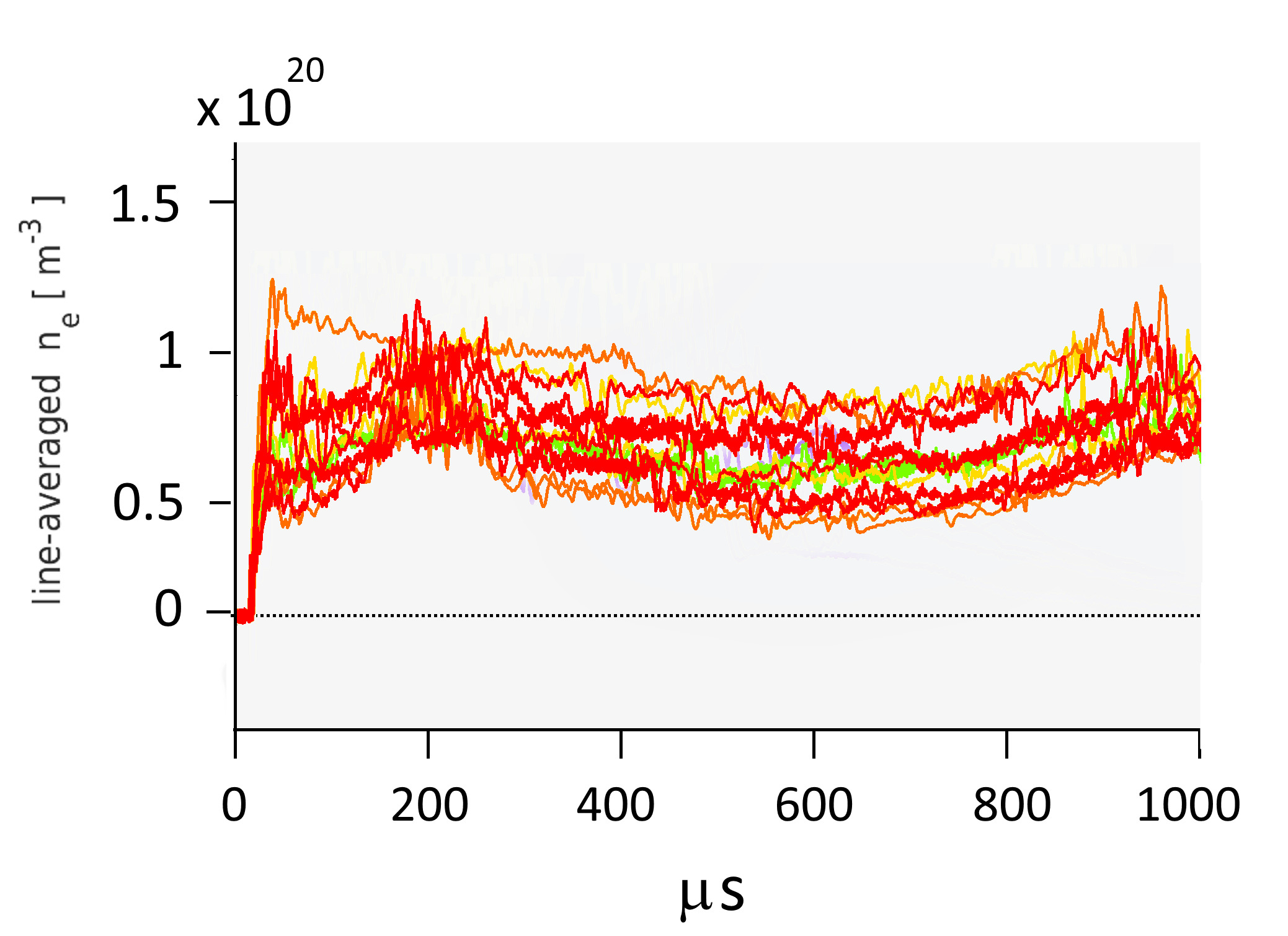}}}\textcolor{black}{\hfill{}}\subfloat[]{\raggedright{}\textcolor{black}{\includegraphics[width=7cm,height=5cm]{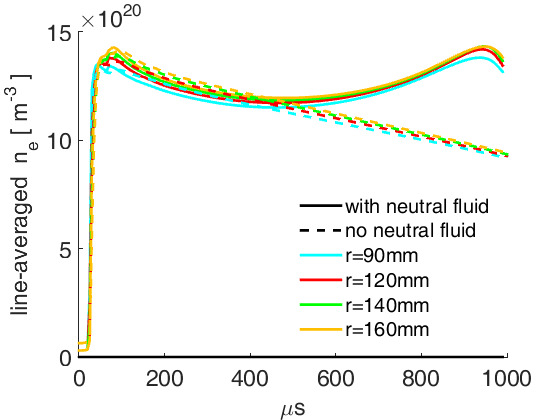}}}

\textcolor{black}{\caption{\label{fig:Spector_nN}$\,\,\,\,$Effect of neutral fluid dynamics
in SPECTOR geometry}
}
\end{figure}
\textcolor{black}{Figure \ref{fig:Spector_nN}(a) shows measured line-averaged
electron density along the chord at $r=140$mm from a selection of
several shots in SPECTOR. It can be seen how density starts to rise
at around 500 to $600\upmu$s. Figure \ref{fig:Spector_nN}(b) shows
the simulated diagnostic for line-averaged electron density along
the chords indicated in figure \ref{fig: Spect_n_Te}(a). The density
rise is qualitatively reproduced when a neutral fluid is included
in the simulation. Similar simulations without the inclusion of neutral
fluid do not indicate this density rise (dashed lines in figure \ref{fig:Spector_nN}(b)).
Note that the simulations presented in figure \ref{fig:Spector_nN}(b)
were run with artificially high plasma density in order to allow for
an increased timestep and moderately short simulation run-times. Hence,
the electron temperatures indicated in figure \ref{fig: Spect_n_Te}(b)
are underestimations of the actual temperatures due to the overestimation
of density in the simulation. The main goal of these simulations was
to demonstrate that the inclusion of neutral fluid interaction can
qualitatively model the observed electron density increase.}

\end{document}